\documentclass[twocolumn]{aastex62}

\usepackage{amsmath}
\usepackage{hyperref}
\usepackage{tikz}
\usepackage{graphicx}
\usepackage[normalem]{ulem}
\usetikzlibrary{positioning}
\usetikzlibrary{patterns}
\usetikzlibrary{decorations.pathmorphing}
\usetikzlibrary{arrows}

\newcommand{\rhogj}{\rho_\mathrm{GJ}}
\newcommand{\gthr}{\gamma_\mathrm{thr}}

\graphicspath{{./}{figures/}}

\shorttitle{Weak Pulsar Magnetospheres}
\shortauthors{Chen et al.}

\begin{document}

\title{Filling the Magnetospheres of Weak Pulsars}

\correspondingauthor{Alexander Y. Chen}
\email{alexc@astro.princeton.edu}

\author{Alexander Y. Chen}
\affil{Department of Astrophysical Sciences, Princeton University, Princeton, NJ 08544, USA}

\author{F\'abio Cruz}
\affiliation{GoLP/Instituto de Plasmas e Fus\~{a}o Nuclear, Instituto Superior T\'ecnico, Universidade de Lisboa, 1049-001 Lisbon, Portugal}

\author{Anatoly Spitkovsky}
\affiliation{Department of Astrophysical Sciences, Princeton University, Princeton, NJ 08544, USA}

\begin{abstract}
    Recent advances in numerical algorithms and computational power have
    enabled first-principles simulations of pulsar magnetospheres using
    Particle-in-Cell (PIC) techniques. These ab-initio simulations seem to
    indicate that pair creation through photon-photon collision at the light
    cylinder is required to sustain the pulsar engine. However, for many
    rotation-powered pulsars pair creation operates effectively only near the
    stellar surface where magnetic field is high. How these ``weak pulsars''
    fill their magnetospheres without efficient photon-photon pair conversion in
    the outer magnetosphere is still an open question. In this paper, we present
    a range of self-consistent solutions to the pulsar magnetosphere that do not
    require pair production near the light cylinder. When pair production is
    very efficient near the star, the pulsar magnetosphere converges to
    previously-reported solutions. However, in the intermediate regime where
    pair supply is barely enough to sustain the magnetospheric current, we
    observe a time-dependent solution with quasi-period about half of the
    rotation period. This new quasi-periodic solution may explain the observed
    pulsar death line without invoking multipolar components near the star, and
    can potentially explain the core vs. conal emission patterns observed in
    pulsar radio signals.
\end{abstract}


\section{Introduction}
\label{sec:intro}

Over the past few years, significant progress towards understanding how pulsars
work has been made with the help of direct Particle-in-Cell (PIC) simulations.
With enough pair supply, the pulsar magnetosphere is well described by the
force-free solution, forming a Y-point near the light cylinder, which connects
an equatorial current sheet and two curved current sheets along the separatrix
between closed and open field lines. Particles are accelerated along these
current sheets, and it is possible to construct the observed gamma-ray light
curves from these first principle simulations \citep{2016MNRAS.457.2401C,
  2018ApJ...855...94P}. It was also shown that general relativistic effects,
although small, are important to allow certain regions of the polar cap to
create pairs, enabling radio emission \citep{2015ApJ...815L..19P}.

The electrodynamics of the pulsar magnetosphere is predicated on sufficient
supply of plasma, but less sensitive to where the plasma is produced. Directly
injecting particles everywhere in the magnetosphere
\citep[e.g.][]{2014ApJ...785L..33P}, injecting pairs artificially from the
surface \citep[e.g.][]{2015MNRAS.448..606C,2018ApJ...858...81B}, or
self-consistent pair creation \citep[e.g.][hereafter CB14]{2015ApJ...801L..19P,
  2014ApJ...795L..22C} all seem to give a global picture similar to force-free,
as long as the pair creation rate is large enough.

However, the location of pair production becomes important when pair supply
becomes low and deviation from force-free becomes more evident. CB14 pointed out
that there are two classes of pulsars, and they may have qualitatively different
magnetospheric structures. Type~I pulsars have significant optical depth to
$\gamma$-$\gamma$ pair production near the light cylinder. These are mostly
young and rapidly rotating pulsars like the Crab, and include most known
gamma-ray pulsars. They tend to form a force-free magnetosphere with Y-shaped
current sheets, launching a pulsar wind of high multiplicity pair plasma. Type
II pulsars are those that do not have enough opacity to $\gamma$-$\gamma$
collision, and their main accessible channel to produce $e^{\pm}$ pairs is
through magnetic conversion. Since this process requires very high magnetic
fields, comparable to the quantum critical field $B_{Q}$, it is only operational
near the stellar surface. These pulsars make up a significant fraction of known
radio pulsars, and understanding how they operate is an important problem. There
are different conclusions in the literature. CB14 concluded that type II pulsars
require misalignment to be active, otherwise they settle to a charge-separated
solution similar to an ``electrosphere'' \citep[e.g.,][]{1985MNRAS.213P..43K}.
\citet{2015MNRAS.448..606C} were able to find an intermediate aligned rotator
solution that has a thick equatorial current sheet and less spindown power than
the force-free solution, using a low rate of pair injection from the surface of
the star.
These type II pulsars were also called ``weak pulsars'' by
\citet{2015arXiv150305158G}, who obtained a solution that has large vacuum
gaps and can convert up to 50\% of the spindown power to radiation
  \citep{2012arXiv1209.5121G}.

In this paper, we investigate how pulsars of type II, or ``weak pulsars'' (we
will use these two terms interchangeably), support their magnetospheres and
produce observable radiation. In Section~\ref{sec:theory} we discuss the
microphysics of polar cap pair creation and how it maps to PIC simulations, in
order to motivate the parameter regimes used in our runs. In
Section~\ref{sec:setup} we discuss our numerical setup and several
implementation choices, and in Section~\ref{sec:results} we present the
solutions we find in different parameter regimes, analyzing them in detail. In
Section~\ref{sec:speculation} we discuss the relevance of the models found
in this paper in the context of existing pulsar theory and phenomenology.
Finally in Section~\ref{sec:discussion} we conclude with a discussion on the
limitations of this paper and possible future directions.

\section{Theoretical Motivation}
\label{sec:theory}

We consider only magnetic conversion of $\gamma$-ray photons into pairs since it
is the dominant pair production mechanism in the magnetospheres of weak pulsars.
The cross-section of this process depends exponentially on the ratio between
  local magnetic field $B$ and the quantum critical field, $B/B_Q$, where
$B_Q = m_{e}^2c^3/\hbar e \approx 4.4\times 10^{13}\,\mathrm{G}$ \citep[see,
e.g.,][]{1966RvMP...38..626E}. When $B \ll B_Q$ this process is exponentially
suppressed, which means it can only operate very close to the star where
$B/B_{Q}\gtrsim 0.1$. It is possible to model this as a sharp cutoff radius
$R_\mathrm{cut}$, outside which pair creation is not allowed. Depending on the
local $B$ field, $R_\mathrm{cut}$ is typically several stellar radii, $R_{*}$.

There are three energy scales governing the pair creation process near the polar
caps of pulsars. The maximum potential drop across a pulsar polar cap can be
estimated as $\Phi_\mathrm{pc} \sim \mu_{B}\Omega^{2}/c^{2}$,
\citep[e.g.,][]{1975ApJ...196...51R}, which translates to a maximum Lorentz
  factor achievable by particles,
\begin{equation}
    \gamma_\mathrm{pc} = \frac{e\Phi_\mathrm{pc}}{m_{e}c^2} \sim 1.2\times 10^{7}B_{12}P^{-2},
    \label{eq:gamma-max}
\end{equation}
assuming $B_0=B_{12}\times 10^{12}\,\mathrm{G}$ at the pole, and period $P$ is
measured in seconds. $\Phi_\mathrm{pc}$ is the maximum potential
drop across the polar cap, but in reality
the polar cap gap may not reach this potential drop, as the gap will
start to be screened as soon as pairs are produced. The actual polar cap
  voltage is expected to be lower and regulated by pair creation activity.

The second energy scale is the energy of accelerated particles that
are capable of emitting pair-producing photons. We loosely call this
pair-creating Lorentz factor $\gthr$. In classic pulsar theory, high energy
gamma-ray photons are produced through the curvature radiation of primary particles,
and the typical photon energy is $\hbar\omega_{c}=3\gamma^{3}\hbar c/2\rho_{c}$,
where $\rho_{c}$ is the radius of curvature of the particle trajectory. The
optical depth of the curvature photons depends on the quantum parameter $\chi \sim
\epsilon_{\gamma}b\sin\psi$, where $\epsilon_{\gamma}$ is the photon energy in
units of $m_{e}c^{2}$, $b = B/B_{Q}$, and $\psi$ is the angle between the photon
momentum and local $B$ field. When the photon is emitted, the angle $\psi$
is negligible due to strong Lorentz boost along the parallel direction, and
$\psi$ builds up nearly linearly with distance traveled by the photon.
Conversion of the photon to pairs happens roughly when $\chi \sim 0.1$
\citep{2019ApJ...871...12T}. If one requires the photon to convert within 1
$R_{*}$ from the surface (before the field strength drops too low), then one can
estimate the pair creation threshold to be:
\begin{equation}
    \label{eq:gamma-thr}
    \gthr \sim \left( \frac{B_{Q}\rho_{c}^2m_{e}c}{15B_{0} R_{*}\hbar} \right)^{1/3} \sim 8.6\times 10^{6}B_{12}^{-1/3}P^{1/3},
\end{equation}
assuming dipole field and curvature radius along the last closed field line.
Depending on the actual $B$ field strength near the surface and field line
geometry this threshold can vary significantly. For example, a multipolar
component near the polar cap can vastly reduce the curvature radius of the field
lines, lowering the threshold by orders of magnitude. The ratio
$\gamma_\mathrm{pc}/\gamma_\mathrm{thr}$ determines whether pair creation is
efficient, and large values of this ratio indicate the ease of converting
  the voltage drop to high pair multiplicity. Conversely, values of
  $\gamma_\mathrm{pc}/\gamma_\mathrm{thr}$ that approach unity indicate
  inefficient pair production, which is associated with the cessation of pulsar
  activity, and corresponds to the pulsar ``death line'' in the $P$-$\dot{P}$
  plane. We will continue the discussion of the pulsar death line in
Section~\ref{sec:speculation}.

\begin{figure*}[t]
    \centering
    \includegraphics[width=0.95\textwidth]{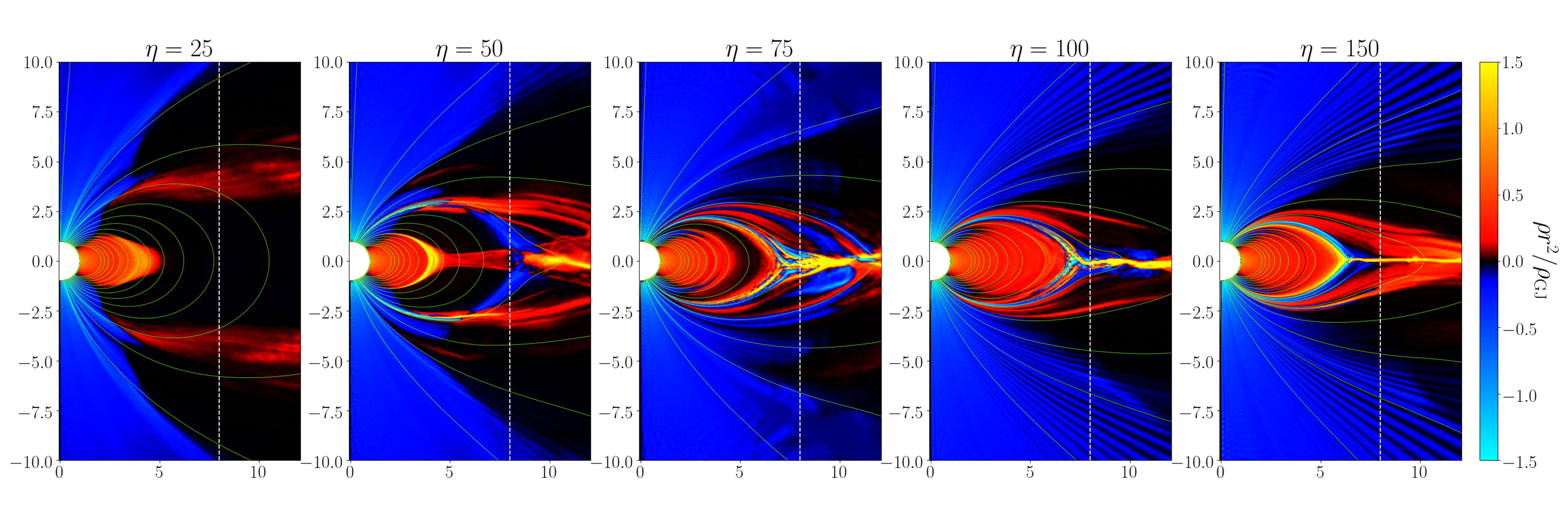}
    \caption{Comparison of 5 different simulations of different ratios
      $\eta = \gamma_\mathrm{pc}/\gamma_\mathrm{thr}$ at the same snapshot
        time $t = 90R_{*}/c$. Plotted in color is the charge density in the
      magnetosphere, multiplied by a factor of $r^{2}$ to better visualize
      features away from the star. Each case is normalized to its own
      $\rho_\mathrm{GJ} = \Omega B_{0}/2\pi c$. Vertical white dashed line
      is the light cylinder, and green curves are magnetic field lines.}
    \label{fig:comparison}
\end{figure*}

The third energy scale is the typical energy of secondary pairs $\gamma_{s}$.
The energies of the curvature photons emitted by the primary particles near
$\gthr$ are much lower than the energy of the emitting particle, and will set an
energy scale for secondary pairs:
\begin{equation}
    \label{eq:gamma-s}
    \gamma_s \sim \frac{3}{4}\frac{\hbar}{m_{e}c^2}\frac{\gamma_\mathrm{thr}^{3}}{\rho_{c}} \sim 200 B_{12}^{-1}P^{-1/2}.
\end{equation}
The ratio $\gamma_\mathrm{thr}/\gamma_{s}$ is a major factor in regulating the
pair multiplicity from the polar cap cascade, as it determines how many
  pairs can one primary particle generate.

In general, for pulsars away from the death line, the energy scales should obey
the hierarchy: $1 \ll\gamma_{s}\ll \gamma_\mathrm{thr} \ll \gamma_\mathrm{pc}$.
In a PIC simulation, especially a global one, this kind of scale separation is
typically not achievable, and one has to rescale the energies preserving the
ordering of scales. For example, in the Type-II case presented in CB14,
$\gamma_\mathrm{pc} \sim 425$ (referred to as $\gamma_{0}$ in that paper),
$\gamma_\mathrm{thr} \sim 25$, and $\gamma_{s} \sim 10$. In the global
simulations published so far, typically $\gamma_\mathrm{pc} \lesssim 1000$, with
$\gamma_\mathrm{thr}$ and $\gamma_{s}$ similar to CB14. This reduced scale
separation has two effects. First, it severely limits the multiplicity of any
pair cascade triggered in the magnetosphere from energetics alone, as a
  primary particle with Lorentz factor $\gamma_\mathrm{thr}$ can only convert
  its energy to one or two $e^{\pm}$ pairs with $\gamma_s$, whereas in reality
  it would be~$\gtrsim 10^3$. It also places the simulated pulsars dangerously
close to the death line, as the ratio $\gamma_\mathrm{pc}/\gamma_\mathrm{thr}$ is
merely of order $\sim 10$.

In this paper, we approach this issue by pushing up the ratio
$\gamma_\mathrm{pc}/\gamma_\mathrm{thr}$, in order to better approximate a
pulsar that is far from the death line but still not energetic enough to
produce pairs near the light cylinder. This is also a regime that is easier
to simulate since when $\gamma_\mathrm{pc}/\gamma_\mathrm{thr} \gg 1$, typical
curvature photons have relatively short free path in the strong magnetic field,
and ``on-the-spot" pair creation scheme is applicable. We vary this ratio to
study the transition of an active rotation-powered radio pulsar to its
death. On the other hand, we keep the ratio $\gamma_\mathrm{thr}/\gamma_{s}$
  low across the simulations in order to keep the pair multiplicity and total number of particles manageable in our
  simulations.

\section{Simulation Setup}
\label{sec:setup}

We simulate an aligned rotator whose magnetic axis is parallel to the rotation
axis, using the code \emph{Aperture} \citep{chen-thesis}. The neutron star is
placed at the origin in log-spherical coordinates. We assume axisymmetry and
simulate the magnetosphere in the $r$-$\theta$ plane. The simulation domain
extends from the stellar surface to about $4R_\mathrm{LC}$, where
$R_\mathrm{LC}$ is the light cylinder radius. Unless stated otherwise, we use
$R_\mathrm{LC}/R_{*} = 8$ and allow pair creation up to radius $R_\mathrm{cut} =
3R_{*}$. Pair creation happens whenever an electron/positron reaches the Lorentz
factor $\gamma_\mathrm{thr} = 25$ within the pair creation radius, and an
$e^{\pm}$ pair is created instantly at $\gamma_{s} = 8$. The simulations shown
in this paper all have resolution $2048\times 2048$, which translates to
  about 650 grid points per $R_{*}$ at the stellar surface.

The star is initially at rest and spins up to the target angular velocity
$\Omega_{*}$ over $t_\mathrm{spin} = 10R_{*}/c$. We start with a pure dipole
  magnetic field in vacuum. Particles injected at the surface are assigned a
weight $w \propto \sin\theta$ which varies with the volume of the cell where they
are injected. This weight carries over to new pairs, and can be understood as
the amount of physical particles represented by a given macro-particle in the
simulation. We apply the spin as a boundary condition at the stellar surface,
$E_{\theta} = -v_{\phi}B_r$, where $v_{\phi}$ is given by the Lense-Thirring
reduced angular velocity \citep{2015ApJ...801L..19P}:
\begin{equation}
\label{eq:v-phi}
v_{\phi} = \frac{(\Omega_{*} - \omega_\mathrm{LT})r\sin\theta}{c\alpha}.
\end{equation}
We use the compactness parameter $r_s/R_{*} = 0.5$ for all our runs. GR effects
are taken into account in the field equations similarly to what was used by
\citet{2018ApJ...855...94P}, namely using a dipole background $B$
field for the $\nabla\times B$ term in the $E$ field update equation. This
GR correction effectively reduces the background charge density $\rhogj = \Omega B/2\pi c$ near the polar cap, increasing the
ratio the current $j$ and $\rhogj c$ to above unity, therefore allowing pair production to happen near the pole.

We also include strong synchrotron cooling to reduce magnetic bottling effect
for plasma flowing towards the star. We damp directly the perpendicular momentum
$p_{\perp}$ of electrons and positrons at every timestep, and the strength of
this damping is directly proportional to local $B$ field. An electron would
typically lose almost all its perpendicular momentum in about 20 timesteps near
the stellar surface.

We define the ratio $\eta = \gamma_\mathrm{pc}/\gamma_\mathrm{thr}$ and fix
$\Omega$ and $\gamma_\mathrm{thr}$, while increasing $\eta$ by increasing
$B_{0}$, which is the dipolar magnetic field strength at the equator of the
star. In the following section, we report the results of this series of
numerical experiments.

\section{Results}
\label{sec:results}

\subsection{A Range of Weak Pulsar Solutions}

By increasing the ratio $\eta$, we see a transition through very different
solutions to the weak pulsar magnetosphere. Figure \ref{fig:comparison} shows a
comparison of 5 runs with increasing $\eta$. The case with $\eta = 25$ is very
similar to the original Type-II solution reported in CB14. The pulsar settles
down to a state with very low spindown power, with a large vacuum gap outside the
pair creation radius $R_\mathrm{cut}$. There is still a small current escaping
the polar cap and along the field line that touches the light cylinder, and some
remnant pair creation activity launches positrons into the vacuum gap which
supports the return current. However, all field lines remain closed, and the
  small remnant current is conducted by escaping charges accelerated by the
  vacuum gap, moving across field lines. On a much longer time scale, we expect
  the current to gradually decrease as the magnetospheric solution settles down to an
  electrosphere.

The cases with $\eta = 50$ and $\eta = 75$ are highly variable. Figure
\ref{fig:comparison} is a snapshot of the magnetospheric configuration, but
both solutions are actually cyclic, with episodes of pair creation along the
return current sheet that launch $e^{\pm}$ pairs towards the light cylinder.
When this quasi-neutral plasma outflows, it screens the electric field up to the
Y-point, and forms a charge cloud there. This charge cloud later depletes,
with electrons flowing back towards the star along the separatrix, and positrons
flowing to infinity in the equatorial current sheet. This case will be discussed in more detail in section~\ref{sec:time-dependent}.

The cases with $\eta = 100$ and $\eta = 150$ are quasi-steady again, with
field lines that go through the light cylinder opening up, forming a stable
Y-point near the light cylinder and persistent return current sheets. We observe
reconnection of the poloidal magnetic flux and plasmoids forming periodically in
the equatorial current sheet near the Y-point, similar to the Type I pulsar
reported in CB14. Both cases approach the force-free limit with similar
polar cap outflow multiplicity.

\begin{figure}[h]
    \centering
    \includegraphics[width=0.48\textwidth]{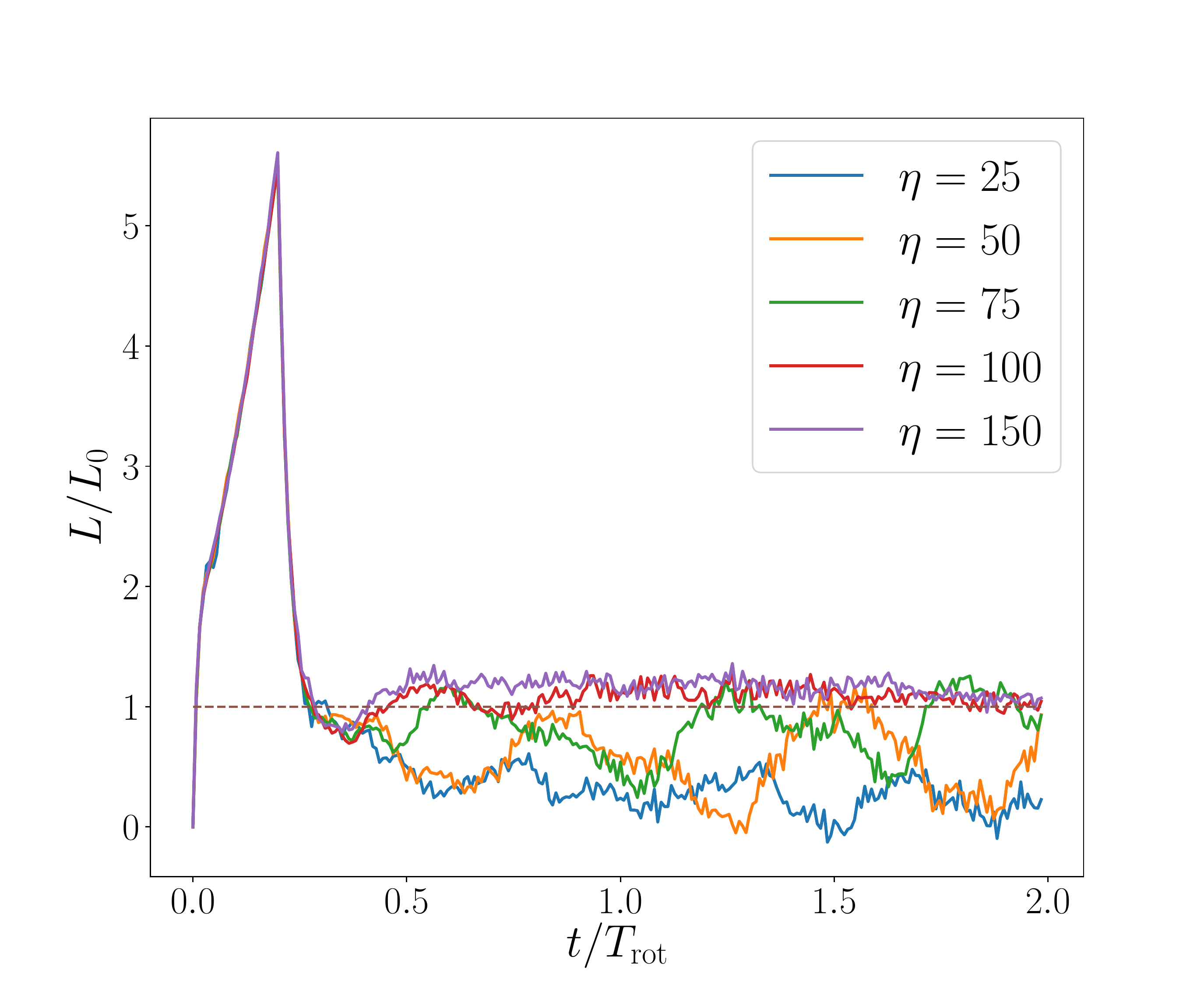}
    \caption{Time evolution of the total Poynting flux from the star for the
        different parameters, normalized to the force-free spindown power $L_0 = \mu^2\Omega^4/c^3$.}
    \label{fig:spindown}
\end{figure}

Figure \ref{fig:spindown} shows the time evolution of spindown power for
the different $\gamma_\mathrm{pc}/\gamma_\mathrm{thr}$ cases, measured as the
integrated Poynting flux from the stellar surface, normalized to the respective
force-free spindown $L_{0} = \mu^{2}\Omega^{4}/c^{3}$ of each run. The Poynting
flux is defined with GR effect taken into account.
It can be seen that after a common initial transient, the runs with $\eta = 100$
and $\eta = 150$ settle to a quasi-steady state with almost force-free
spindown, whereas the intermediate $\eta$ runs show quasi-periodic swings in
spindown luminosity which are in-phase with the pair creation episodes. The
$\eta = 25$ case sees a gradual drop in spindown luminosity, approaching
roughly $0.1$--$0.2$ of the force-free spindown. We expect the spindown
  power to slowly decrease to zero, as the magnetosphere settles down to a state
  similar to the electrosphere. This was the fate of the weak pulsar proposed in
  CB14.

\begin{figure}[h]
    \centering
    \includegraphics[width=0.48\textwidth]{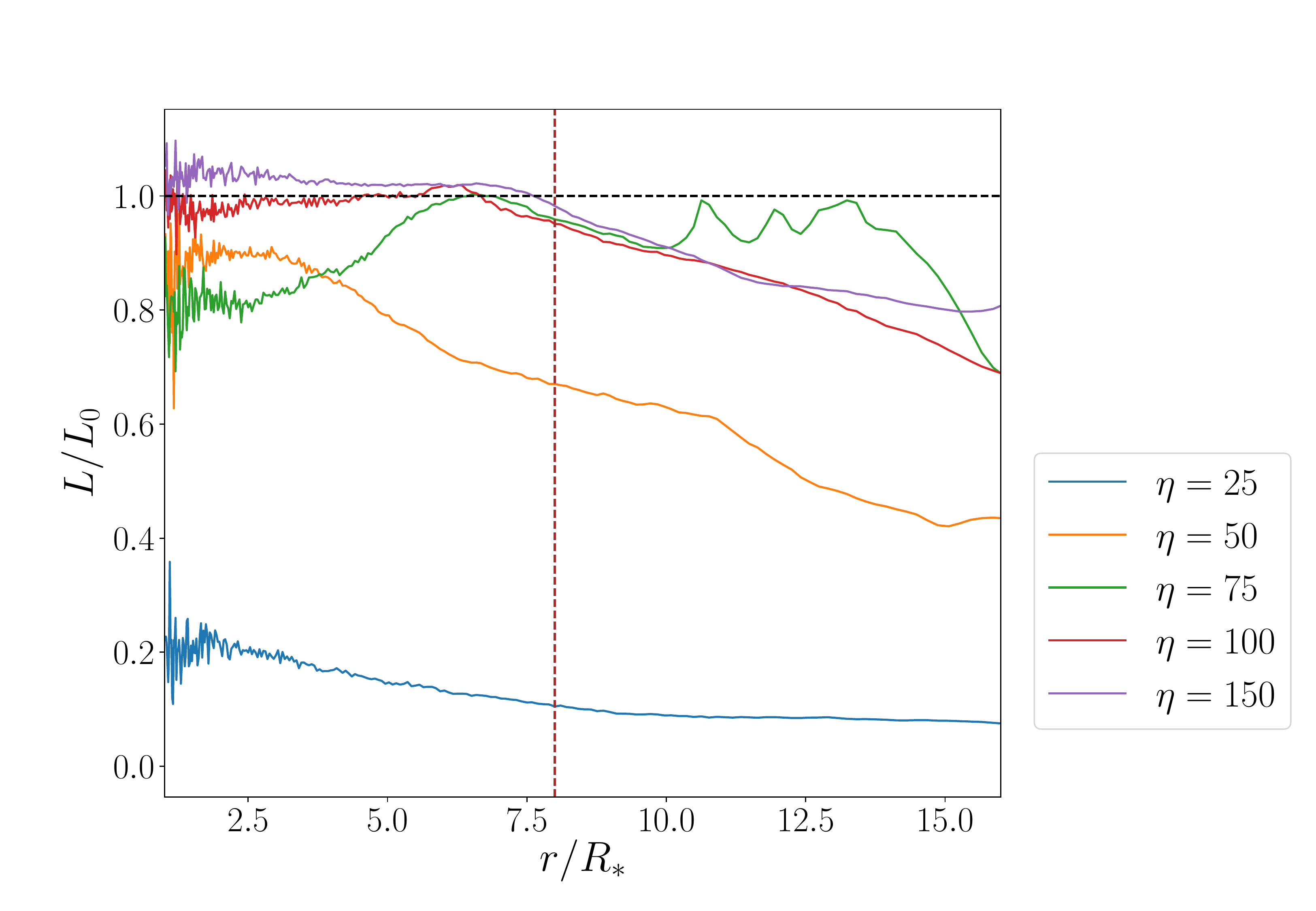}
    \caption{Integrated Poynting flux as a function of radius, for the
        different cases discussed in the text, all normalized to the force-free
        spindown $L_0 = \mu^2\Omega^4/c^3$. This snapshot is taken at $t \sim
        2T_\mathrm{rot}$. Vertical dashed line marks the light cylinder.}
    \label{fig:dissipation}
\end{figure}

Figure~\ref{fig:dissipation} shows the radial dependence of the integrated
Poynting flux for the different cases. It can be seen that $\eta = 100$ and
$\eta = 150$ cases have very little dissipation of the Poynting flux inside the
light cylinder, but about 20\% of it is dissipated between $r=R_\mathrm{LC}$ and
$2R_\mathrm{LC}$. The dissipation mainly happens in the equatorial current
sheet. This agrees very well quantitatively with the high particle injection
case reported by \citet{2015MNRAS.448..606C}. As a result, we expect most of the
high energy $\gamma$-ray emission for these pulsars to come from outside the
light cylinder, in the equatorial current sheet, similar to what was reported by
\citet{2016MNRAS.457.2401C} and \citet{2018ApJ...855...94P}. The intermediate
solutions, however, see a somewhat higher dissipation inside the light cylinder,
up to 10\%--20\%. This suggests that in these solutions a fraction of the
spindown power can in principle be dissipated into particle kinetic energy
within the light cylinder, in agreement with the recurrent large vacuum
gaps reported in section~\ref{sec:time-dependent}. In
Figure~\ref{fig:dissipation}, the $\eta=75$ case seems to suggest energy
injection near the light cylinder, but it is simply a result of the
time-dependent nature of the solution, as the Poynting flux from the star goes
through large-amplitude oscillations.

\begin{figure}[h]
    \centering
    \includegraphics[width=0.46\textwidth]{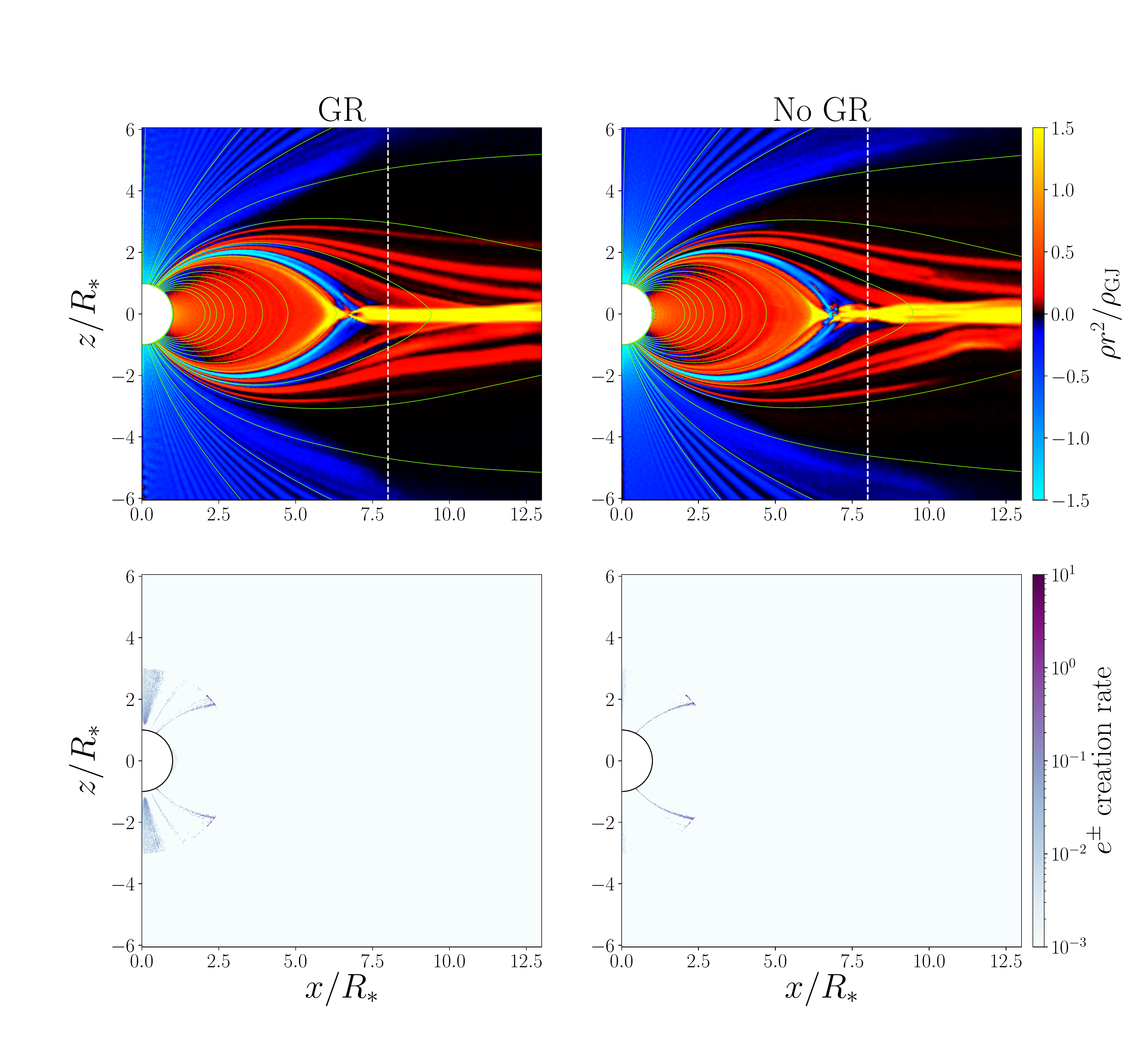}
    \caption{Pair creation sites, with and without GR, in the $\eta=100$ case.
      Snapshot is taken at $t \approx 2T_\mathrm{rot}$. The upper panels show
      the charge density and magnetic field structure similar to
      Figure~\ref{fig:comparison}, while the lower panels show pair creation
      rates per cell normalized to arbitrary units around the same time, averaged over $\Delta T = R_*/c$. Pair
      creation at the center of the polar cap is notably missing in the non-GR
      simulation, while both have pair creation along the separatrix current
      sheet. Even without polar cap pair production, the magnetospheric
      structure remains exactly the same.}\label{fig:pair-creation-sites}
\end{figure}

It is also instructive to look at where pairs are created.
Figure~\ref{fig:pair-creation-sites} shows the pair creation sites, with and
without GR effect. It can be seen that pairs are only created at the center of
the polar cap and along the separatrix current sheet. These are the sites
  where $j/\rhogj c$ is either above unity or below zero, which are expected to
  require pair production to conduct the magnetospheric current \citep{2008ApJ...683L..41B}. Without GR effect, there is
no polar cap pair creation, in line with what was reported by
\citet{2015ApJ...815L..19P}. It is curious, however, that even without GR effect
and pair creation at the center of the polar cap, the pulsar is capable of
supporting the structure of the magnetosphere with only pairs created along the
separatrix current sheet. If we believe that radio emission is associated with high
plasma multiplicity at the polar cap, then GR effect is essential for turning on
radio emission for many pulsars, especially if their rotation and magnetic axes
are nearly aligned. However, it turns out to be not so important for the structure
of the magnetosphere, as the magnetospheric current can be conducted simply by
extracted electrons flowing at mildly relativistic speeds, which agrees with
what was reported by \citet{2013ApJ...762...76C} and
\citet{2013MNRAS.429...20T}.

\begin{figure}[h]
    \centering
    \includegraphics[width=0.45\textwidth]{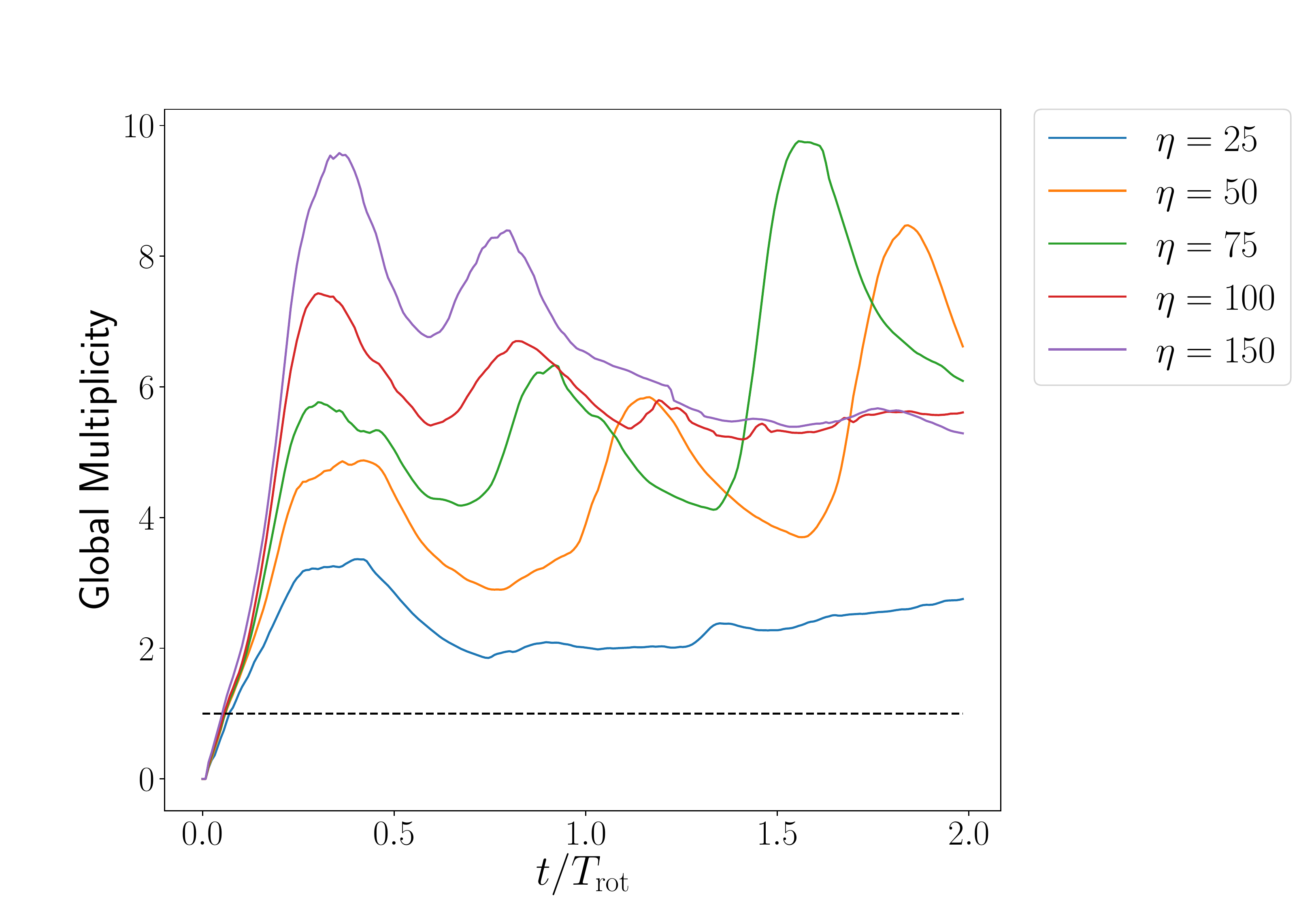}
    \caption{Global multiplicity evolution for the near force-free cases. Black dashed line represents $\mathcal{M} = 1$. The higher $\eta$ runs have higher multiplicity during the initial transient. The $\eta=100$ and 150 cases both relax to a similar global multiplicity, while the two intermediate runs show the same kind of quasi-periodic oscillations in the multiplicity. The lowest $\eta$ run is never able to sustain the same amount of pairs, as expected from the spindown comparison.}
    \label{fig:multiplicity}
\end{figure}

Plasma supply is the key factor that differentiates the range of pulsar
solutions. One way to quantify this is to define the global pair multiplicity,
defined as:
\begin{equation}
    \label{eq:global-M}
    \displaystyle \mathcal{M} = \frac{\int\,d\Omega\,\int_{R_{*}}^{R_\mathrm{LC}}e(n_{+} + n_{-})\,dr}{ \int\,d\Omega\,\int_{R_{*}}^{R_\mathrm{LC}}|\rho_\mathrm{GJ}|\,dr},
\end{equation}
where $n_\pm$ are electron and positron number densities. This quantity measures how
much plasma is produced in excess to the minimum GJ charge density.
Figure~\ref{fig:multiplicity} shows the time evolution of this multiplicity for
the different runs considered. It can be seen that going from $\eta = 100$ to
$\eta = 150$ increases the pair multiplicity during the initial transient, but
the system settles down to a similar global multiplicity. The polar cap
acceleration potential $\Phi$ in these two cases also capped at
$\Phi_\mathrm{thr}$, much less than the theoretical $\Phi_\mathrm{pc}$, in
agreement with the discussion in Section~\ref{sec:theory}. The intermediate
cases see oscillations in multiplicity that mirror the time-dependence in the
light curves. The fact that the overall global multiplicity is increasing for
both cases strongly suggests that these solutions are self-sustaining and should
be stable in the long term.

\subsection{The Oscillatory Intermediate Solution}
\label{sec:time-dependent}

\begin{figure*}[t]
    \centering
    \includegraphics[width=0.95\textwidth]{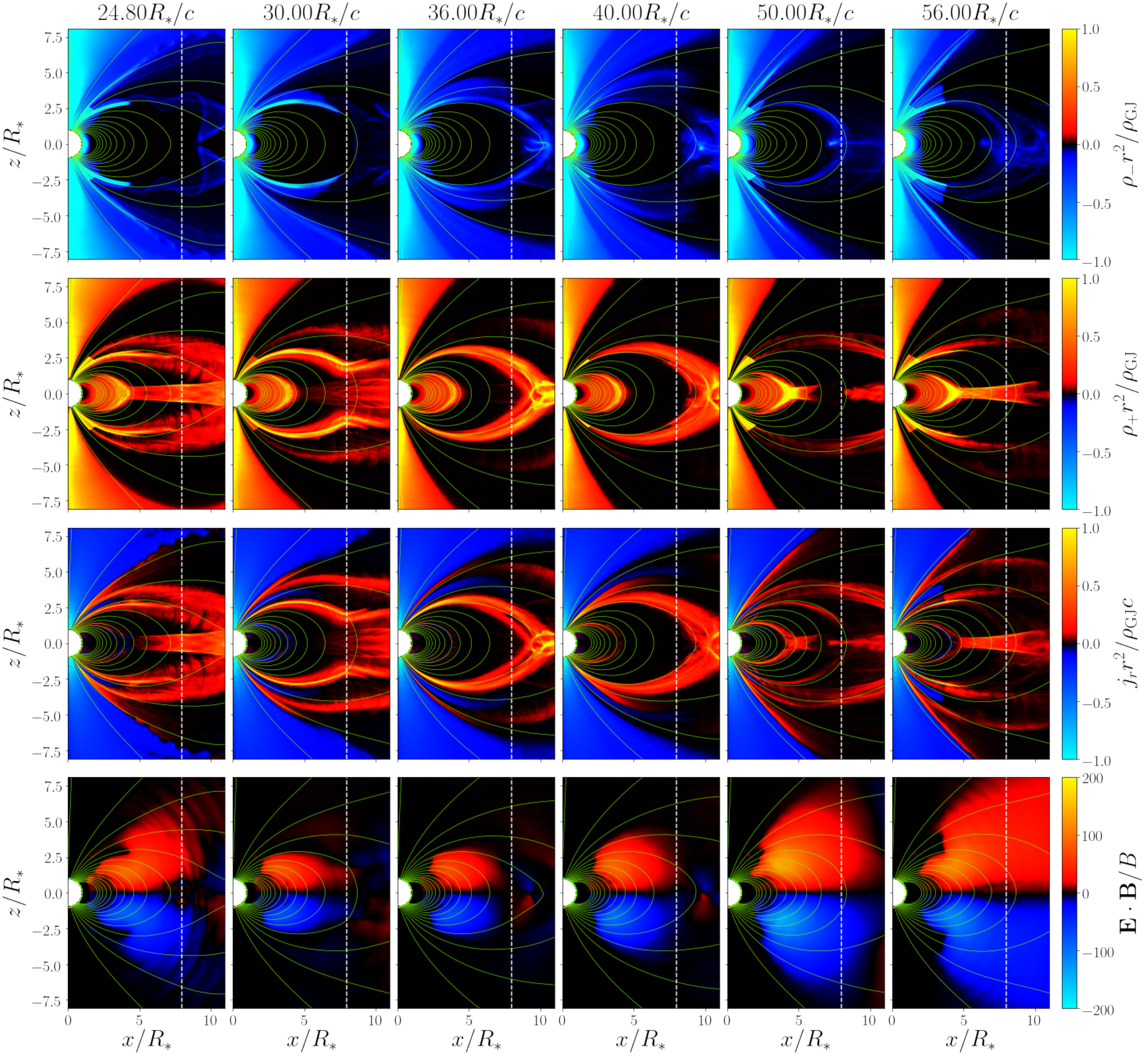}
    \caption{The $\eta = 50$ case in detail. From left to right shows evolution
      in time. The six panels together show a complete cycle. From top to bottom
      the color plots show a) electron density $\rho_{-}r^2$, b) positron
      density $\rho_+r^2$, c) radial current $j_{r}r^2$, d) $\mathbf{E}\cdot \mathbf{B}/B$. The charge
      and current densities are normalized to $\rhogj$ at the pole, and $\mathbf{E}\cdot
      \mathbf{B}$ is in numerical units. White vertical dashed lines mark the
      light cylinder, and green lines show the magnetic field.}\label{fig:time-dependent}
\end{figure*}

The intermediate solution where the magnetospheric structure of the pulsar is
highly time-dependent is particularly interesting.
Figure~\ref{fig:time-dependent} shows the time evolution of various
magnetospheric quantities over one such cycle. It starts with quasi-neutral
plasma outflowing from the pair creation surface at $r \sim 3R_{*}$ (column 1,
electron and positron densities near the separatrix), screening the parallel
electric field along the separatrix\footnote{The ``separatrix'', usually
  denoting the return current sheet formed along the last closed field line,
  loses the usual meaning here as most field lines remain closed. We simply use
  this term to denote the strong return current flowing in the vicinity of the
  usual separatrix.}, momentarily forming the separatrix current sheet and the
Y-point outside the light cylinder (columns 2-3, see current and
$\mathbf{E}\cdot \mathbf{B}$). As the parallel electric field becomes screened,
pair creation activity is reduced, and the return current can no longer be
sustained. At this point the electrons in the outer magnetosphere start to fall
back onto the star to conduct the return current (columns 4-5, see electron
density). These electrons are accelerated towards the star because
$E_{\parallel}$ is induced again due to insufficient current. They start to
create pairs when they reach the pair creation surface at $r \sim 3R_{*}$ and
reignite the pair creation, eventually launching a quasi-neutral outflow again,
initiating the next cycle (columns 5-6, electron and positron densities).

Even in the quasi-stationary force-free like solution with $\eta = 100$ or
$150$, we still observe this cyclic behavior to some degree. The pairs
outflowing along the separatrix current sheet tend to form a charge cloud near
the Y-point. A stream of electrons is drawn from the Y-point cloud to help
support the return current, and these electrons are accelerated on their way
towards the star, producing pairs as they come close to the stellar
  surface. A fraction of these pairs outflow again, carrying stronger return
current and screening $E_{\parallel}$ along the separatrix, thus reducing the
acceleration voltage on the returning electrons and suppressing pair creation. The
$E_{\parallel}$ oscillations are much smaller in amplitude than in the
intermediate regime, but could potentially contribute to the complex
time-dependent behavior we see in pulsar radio emission.

We observe the period of this quasi-cycle to be near half of the rotation
period, comparable to the travel time between the star and the light cylinder
for relativistic particles along the last closed field line. However we believe
the period can depend on the multiplicity from the pair cascade near $r\sim
3R_{*}$, since if more pairs outflow during the active phase, more electrons can
be stored in the Y-point charge cloud which takes longer to deplete. If this is
true, higher multiplicity should translate to a longer duty cycle. In this paper
we are unable to perform simulations with much higher multiplicity due to
computational constraints, since higher multiplicity require a finer grid,
consumes larger memory, and leads to high local concentration of particles that
make simulations difficult. We will defer the study of the multiplicity
dependence of the cyclic solution to a future work.

\section{Implications of the Time-Dependent Weak Pulsar Model}
\label{sec:speculation}

The time-dependent intermediate solution presented in
Section~\ref{sec:time-dependent} is a result of
$\gamma_\mathrm{pc}/\gamma_\mathrm{thr} = 50$, but is more representative of
pulsars when the polar cap pair supply is marginally enough to sustain the
magnetosphere. The same description is usually applied to pulsars near the death
line, as it is conventionally believed that a high plasma multiplicity from pair
creation is essential in producing the observed pulsar radio emission. The
pulsar death line has been studied in great detail before
\citep[e.g.][]{1993ApJ...402..264C, 2000ApJ...531L.135Z, 2001ApJ...554..624H},
and most authors conclude that a dipolar magnetic field is not enough to explain
the observed death line. In fact, if the high energy $\gamma$ rays from
curvature radiation are the main pair-creating photons, then the death line
computed from polar cap voltage assuming a dipole field configuration can be
found by equating $\gamma_\mathrm{pc}$ to $\gamma_\mathrm{thr}$
(equations~\eqref{eq:gamma-max} and \eqref{eq:gamma-thr}), which lies somewhere
in the middle of the observed pulsar population \citep{1993ApJ...402..264C}.
Usually some non-dipolar configuration is invoked to decrease the radius of
  curvature of pair-producing field lines, and push the death line down to
allow for many observed weak pulsars, regardless of whether curvature radiation
or inverse Compton scattering is the main $\gamma$ ray producing mechanism.

Another seemingly unrelated piece of the puzzle is that isolated
  rotation-powered radio pulsars fall in two populations. Younger and
more energetic pulsars tend to have a clearly defined core radio emission
pattern, while the older and less energetic pulsars tend to have a multiple
conal emission structure \citep[see, e.g.,][]{1983ApJ...274..333R}. The line
separating these two populations is surprisingly close to the naive death line
of a dipole polar cap cascade model \citep[see, e.g.,][]{1997ApJ...474..407W}. The
details of pair creation mechanism and the assumptions in the model
(curvature vs ICS, vacuum gap vs slot gap) may shift the line up or down, but
not by much. It seems a contrived coincidence that the conal emission mainly
comes from pulsars that require some pair creation mechanism that is beyond the
simple dipole model.

In light of our new model for weak pulsars, we propose a potential solution to
both these puzzles that does not require non-dipolar field configuration. As
discussed in Section~\ref{sec:time-dependent}, when pair supply from the star is
not enough, a vacuum gap is periodically opened around the separatrix up to the
light cylinder. This gap, accelerating electrons back towards the star, can
reach much higher potential drop than the maximum polar cap potential
$\Phi_\mathrm{pc} \sim \mu_B\Omega^2/c^2$, approaching the vacuum potential drop
$\Phi_\mathrm{0} \sim \mu_B \Omega/R_*c$. This makes pair creation possible even
if the polar cap voltage is insufficient, but the pair creation activity will be
confined to the return current, along a ring-like structure around the
polar cap. This could in principle lead to the disappearance of the strong core
radio component produced by pairs created at the center of the polar cap, and
the appearance of a conal component near the edge of the polar cap due to the
pairs created along the current sheet.

Pulsars with conal emission often also exhibit the ``drifting subpulses''
phenomenon \citep[see, e.g.,][]{1993ApJ...405..285R}. The timescale for these
subpulse modulations is typically observed to be $P_{3}\sim
2\text{--}15\,\mathrm{period\ cycle^{-1}}$. These modulations were thought to be
associated with drifting local pair discharge activity, or local ``sparks''
\citep[e.g.][]{1975ApJ...196...51R}. A potential alternative explanation could
simply be that $P_3$ is the cycle presented in Section~\ref{sec:time-dependent},
and is regulated by the plasma flow between the stellar surface and the pair
cloud near the light cylinder, especially if the period for the cyclic behavior
scales with the pair multiplicity. In order to validate or disprove this
hypothesis, 3D simulations in the similar parameter regime is likely needed.


\section{Discussion}
\label{sec:discussion}

In this paper we presented a range of self-consistent solutions of the pulsar
magnetosphere. We found that even when pair creation is restricted to be near
the surface, weak pulsars can produce enough pairs to fill the magnetosphere
and reach a near force-free state.

Depending on how easy it is to produce the pairs, these weak pulsars may
settle down to a near-death state as reported in CB14, or stay in a highly
variable state where pair creation and current flow are intermittent, or reach a
near force-free state that is very similar to Type~I pulsars. This should be
compared with the range of solutions obtained by \citet{2015MNRAS.448..606C}
where pairs are supplied artificially from the stellar surface. What we found in
this paper is that with copious pair supply near the star ($\eta \gtrsim
100$), our result with self-consistent pair creation is indeed very similar to
the high pair injection rate solution reported by
  \citet{2015MNRAS.448..606C}: the magnetosphere is near force-free inside the
light cylinder, and most of the Poynting flux dissipation happens outside the
Y-point. However, in the low pair supply regime steady pair creation near
the surface is not possible, and the magnetosphere has to go through episodes of
opening and screening of the pair-accelerating gap, significantly increasing the
dissipation inside the magnetosphere.

Compared with the results obtained by \citet{2013arXiv1303.4094G}, our low to
intermediate pair supply ($\eta \lesssim 75$) solutions are somewhat similar in
the sense that a large unscreened gap can exist in the outer magnetosphere.
Especially in our $\eta = 25$, we do see a small amount of positrons flowing out
near the separatrix, accelerated by the vacuum gap, and move across magnetic
field lines. However, this solution has very low spindown power in the first
place, and the pulsar activity is decreasing over time. Moreover, Gruzinov's
solution did not contain a similar time-dependence that we reported in
section~\ref{sec:time-dependent}. In the case of high pair supply, the
magnetospheric differences between strong and weak pulsars virtually vanish, and
we no longer observe such large vacuum gaps as our solutions become almost
force-free.

In this paper we have used quite a simple model for pair creation, namely,
whenever a particle hits a Lorentz factor threshold, it will immediately create
an $e^{\pm}$ pair. In reality, the microphysics is much more complicated, as
curvature photons will have an energy-dependent free path, which will allow
particle acceleration beyond the threshold energy, and synchrotron cascade will
further enhance the pair multiplicity by about an order of magnitude. Higher
pair multiplicity from the cascade will definitely change the critical ratio
$\gamma_\mathrm{pc}/\gamma_\mathrm{thr}$, potentially enabling the peculiar
time-dependent pulsar solution when $\eta$ is closer to unity. However,
particularly when approaching the death line of $\gamma_\mathrm{pc} \sim
\gamma_\mathrm{thr}$, finite photon free path depending on its energy and
propagation direction will likely play a very important role in the global
plasma supply and dynamics. This should be studied in detail in the future
  with a more sophisticated model for pair creation including the cross section for the pair production process derived from Quantum Electrodynamics, similar to what was developed by
  \citet{PhysRevE.95.023210}.

  \acknowledgments We thank Andrei Beloborodov and Alexander Philippov for
  stimulating and helpful discussions. Some of the ideas in this paper were
  inspired by the discussions at the workshop ``Magnetospheres of Neutron Stars
  and Black Holes'' at Goddard Space Flight Center in June 2019. The code
  \emph{Aperture} used in this work can be found at
  https://github.com/fizban007/Aperture3.git. This work was supported by NASA
  grants NNX15AM30G and 80NSSC18K1099 and by the National Science Foundation
  under Grant No. NSF PHY-1748958. We also gratefully acknowledge the support of
  NVIDIA Corporation with the donation of the Quadro P6000 GPU used for this
  research. FC is supported by the European Research Council (grant InPairs
  ERC-2015-AdG 695088) and the Funda\c{c}\~{a}o para a Ci\^{e}ncia e a
  Tecnologia (FCT, grants PD/BD/114307/2016 and APPLAuSE PD/00505/2012). AS is
  supported by Simons Foundation (grant 267233).

\bibliographystyle{aasjournal}

\end{document}